LAL 16-029

# Diphoton resonance at e⁺e⁻ and $\gamma\gamma$ colliders

F. RICHARD[1]

Laboratoire de l'Accélérateur Linéaire,

IN2P3-CNRS et Université Paris-Sud XI, Bât. 200, BP 34, 91898 Orsay Cedex France

Abstract

*In this note, I will review the opportunities offered by the hint of a new resonance observed at LHC for future e+e- TeV linear collider (LC) projects. This discussion is mainly influenced by two specific scenarios of physics which assume either a (pseudo-)scalar or a tensor resonance, but these estimates can be used in most scenarios. I envisage either a photon collider, which has a guaranteed signal with the LHC observation, or a standard e+e- collider, more straightforward to implement. After a detailed study of the heavy graviton scenario, I conclude that at a TeV LC, high accuracy measurements, including rare modes, allow to unambiguously establish the origin of this resonance. Also envisaged in some detail is a radion scenario which illustrates the production of a scalar. The role of an LC for precision measurements on Higgs and top couplings is recalled in the context of the Randall Sundrum model.*

---

[1] richard@lal.in2p3.fr

[2] https://www.linearcollider.org/P-D/Working-groups

[3] https://agenda.linearcollider.org/event/7014/

# Introduction

It goes without saying that the ATLAS+CMS 'evidence', with a recent update in the 'Rencontres de Moriond 2016', [1] and [2], needs confirmation. This will certainly not stop wild speculations, including for what concerns future machines. In this paper, I will study the obvious case for a photon linear collider (PLC) tuned on the new resonance and, less obvious but not excluded, the production of X(750) through e+e- annihilation at 750 GeV centre of mass energy. Another possibility for a LC, also discussed in section II is associated production of this new object but is requires higher energy.

A PLC scenario requires a Linear Collider (LC) reaching 1 TeV which constitutes an important argument for the choice of future machines.

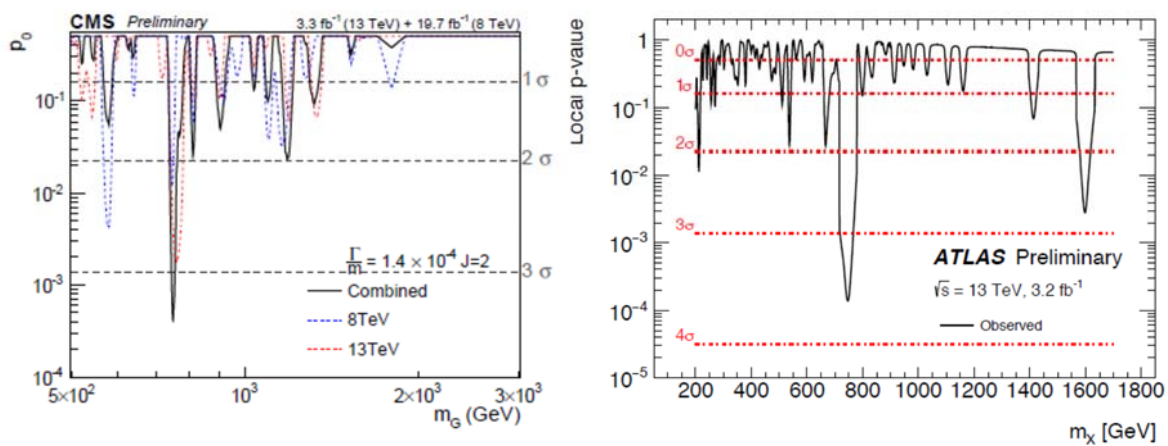

*Figure 1: Last results from ATLAS and CMS on the di-photon anomaly as they were shown in Moriond EW.*

In spite of an avalanche of interpretations[4]: **KK graviton, radion, gauge-higgs unification, dilaton, inflaton, heavy axion, techni-pion, sgolstino, Higgs-like boson (pseudo-scalar, scalar or both), twin Higgs, heavy quark bound states, closed strings, little KK graviton, unparticle, 331 models** etc… very few models, in fact almost none, can claim a prediction. About 300 papers have appeared on this topic and it goes beyond the purpose of this document to attempt review them. While I gratefully recognize my debt to this gigantic effort, only few references directly related to my topic will be quoted. Apologies to the others.

Most interpretations tend to assume that this object has been produced through gluon-gluon fusion. They include the possibility that it is a **scalar or a pseudo-scalar** and, more rarely but with increasing frequency, a **tensor** particle. It will therefore be essential to first determine its spin and then CP. While this was simple in the case of h(125) by using the ZZ* mode, it may be a non-trivial matter in the present case and I will show the unique opportunity offered by a PLC with transversely polarized photons to measure CP.

I have retained two models which claim to have predicted such a particle, being compatible will all existing observables (direct and indirect). These models are not just ad hoc interpretations but also claim to offer solutions to the hierarchy problems, as is the case for SUSY.

The Randall Sundrum (RS) scheme [3] predicts a scalar called the **radion** and, quoting [4], this scalar would be the *lightest object of the theory*, presumably lighter than a TeV and certainly observable at LHC through gluon-gluon fusion. The couplings to SM particles depend crucially on the Ricci-Higgs mixing parameter ξ. [5] gives a detailed interpretation of the LHC observation in terms of a radion assuming that ξ is close to 1/6, a value which corresponds to the so-called conformal limit. Very similar arguments can be drawn concerning the **dilaton** which, quoting [6], could be the **lightest new state** seen at colliders. As pointed out in [6], the dilaton, the radion and the technipion are intertwined concepts in a **composite** description of nature. Section III shows how a PLC could thoroughly study such particles.

---

[4] See http://inspirehep.net/search?ln=en&p=refersto%3Arecid%3A1410174

In the RS model, there is also room for a tensor particle which would correspond to the KK excitation of a **graviton** which also couples to gluon-gluon. Similarly such particles decay into ZZ, WW, hh and top pairs. In this scenario the coupling to e+e- is model dependent but is not excluded. The mass of this resonance is embarrassingly small for usual RS models but, as discussed in section IV, can be accommodated in certain extensions. An LC would then provide an ideal tool to fully study such resonance as demonstrated in section IV.

In the so-called 'walking technicolour' scheme, WTC, the lightest particles are scalars or pseudo-scalars which couple to gluon-gluon and decay into two photons with sufficient rate. [7] gives such an interpretation. It remains to be proven that WTC can accommodate the Higgs measured couplings [8] as claimed by [9]. I will come back to this point.

The so-called 'Twin Higgs' models also predicts a heavy scalar at ~1 TeV, which can be produced in gluon-gluon collisions but, according to [10], this happens with a small cross section, given that the coupling to top quarks is suppressed. It would therefore need an enhanced coupling to photon pairs to account to the LHC observation. While not excluded, given that there are partner particles to W/Z, this possibility has, to my knowledge, not been put forward in the literature.

Not to be forgotten are the various avatars of SUSY which also claim that heavy Higgs bosons could provide an interpretation of this signal with more or less natural assumptions but it seems that other expected manifestations of SUSY are still missing to support such claim. Future results from LHC will tell.

One major goal of our field is to understand the origin of baryon anti-baryon asymmetry in our universe. While the SM alone cannot provide the necessary mechanism to trigger a first order phase transition, adding extra scalars could allow for this. In [11], a workable scheme including X(750) is proposed with two major observable consequences. Firstly, gg→X→2h has a large cross section, above 20 fb at 13 TeV, therefore observable very soon. Second, the Higgs self-coupling constant is increased by 40%, which seems beyond LHC reach but achievable at a TeV e+e- collider. I will return to this point in section V.2.

Another goal of our field is to understand the origin of dark matter in our universe and many papers have developed this aspect. Here I will only briefly discuss this topic in section VII.

Concerning **future machines**, again the inescapable consequence of this signal would be the need for lepton colliders reaching ~1 TeV. I have already pointed out that for a KK graviton, an e+e- collider could provide an ideal tool, easy to implement and only requiring an energy of 750 GeV. This aspect is quantitatively studied in sections II.2.2 and IV.2.

The photon collider scheme, PLC, as already noticed in [12],[13] and [14], will certainly receive increased attention since it can cover both the scalar and the tensor scenarios. It goes beyond the scope of this note to recall the challenges of such a device, but the long standing effort towards a PLC allows to evaluate the rates and the backgrounds which can be expected. Sections III and IV allow to understand these aspects. The importance of polarized photon beams is illustrated on various occasions, in particular in section III.2 for determining the CP properties of this new resonance.

A LC collider will start with a 500 GeV to precisely measure the couplings of the Higgs boson and the top EW couplings. The power of these measurements is recalled since it allows to reach mass scales well above the reach of LHC.

Finally I will go through various related aspects: impact for our universe (section VII), predicted new particles (section VIII), indirect constraints from precision measurements (section IX).

# I General considerations

X(750) is a(pseudo)scalar or a tensor particle, if it decays into two genuine photons. Contrary to a SM Higgs boson, it does not decay copiously into WW/ZZ (for tt the present limit given below is less constraining). For instance, [15] summarizes present relevant searches:

| Final state | ee+μμ | ττ | Zγ | ZZ | Zh | hh | WW | tt | bb | Invis. | jj |
|---|---|---|---|---|---|---|---|---|---|---|---|
| σ at 8 TeV | <1.2 fb | <12 fb | <11 fb | <12 fb | <19 fb | <39 fb | <40 fb | <450 fb | < 1 pb | <0.8 pb | <2.5 pb |

Except for the leptonic pairs, the sensitivity of LHC is rather poor given that in the examples discussed below (heavy graviton and radion) the BR for decays are, with the exception of gg, of same order as the γγ mode. For tt, bb, jj and invisible modes, **detection at LHC seems very problematic**.

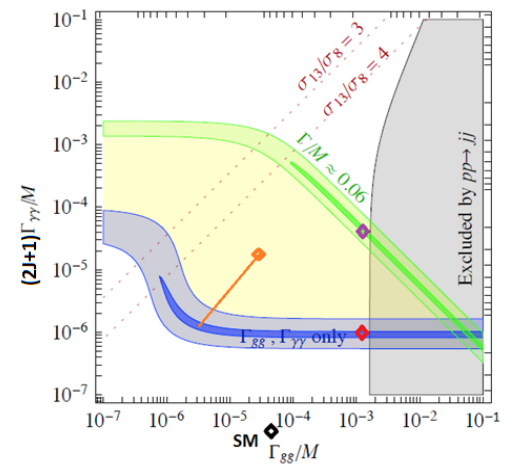

*Figure 2: In this picture, taken from [15], the 3 solutions discussed in the text are displayed by coloured diamonds while a SM Higgs solution would correspond to the black diamond. The various bands and areas are explained in [15]. The orange straight line corresponds to the KK graviton possible solutions with $\Gamma\gamma\gamma$ multiplied by (2J+1) and $\Gamma gg/\Gamma\gamma\gamma=8$.*

Note that the JJ limit, below 1 TeV, comes from a special data taking procedure described in [16].

Assuming that X(750) is produced by gg, the LHC cross section depends on $\Gamma gg\Gamma\gamma\gamma/\Gamma tot$. Hence the constraint given by [15] for a resonance, where M is the mass of the resonance:

$$(2J+1)(\Gamma gg/M)(\Gamma\gamma\gamma/M) \sim 10^{-6}(\Gamma tot/M)$$

where one allows for the possibility of having either a scalar (J=0) or a tensor (J=2) solution. This is only a rough estimate which can be wrong by a factor of ~2.

Given that $\Gamma gg<\Gamma tot$, **(2J+1)Γγγ≥1 MeV**. The **red diamond solution,** shown in figure 2, corresponds to the minimal value which naturally occurs in models with direct couplings like for the radion or graviton solutions in RS or a techni-meson in WTC [17]. It does not come out naturally in extended Higgs models where the coupling to two photons occurs through loops as for the SM Higgs solution for Mh=750 GeV (the black diamond solution). One needs to postulate extra particles enhancing the loop contributions.

Most models fail to reproduce the large total width suggested by ATLAS, **Γtot=45 GeV**. One could assume that there are two nearby Higgs-like resonances almost mass degenerate as suggested in [18] and [19].

In an agnostic approach one can relax these constraints and assume that other modes could substantially contribute to the total width. From the present limits from LHC8 on tt/WW/ZZ/hh/… (or invisible) one could have **tt+hh+…/γγ ~1000**, meaning that allowing for Γγγ =45 MeV one could reach **Γtot~45 GeV**: this is the magenta diamond solution shown in figure 2. The latter value would favour dramatically the scenario of a photon collider as shown in section III.

In section IV, a Kaluza Klein (KK) graviton scenario will be presented and this type of solution corresponds to the orange diamond with a line corresponding to Γgg=8Γγγ. The graviton solution gives a cross section enhanced by the spin factor (2J+1). For consistency, Γγγ is thus replaced by (2J+1)Γγγ in figure 2.

# II How to produce X(750) at e+e- colliders

## II.1 The standard way: e+e-→XZ and Xγ.

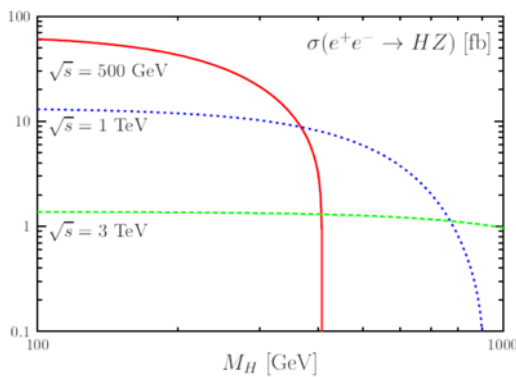

Figure 3: SM model Higgs production cross section for the process e+e-->HZ in fb at 3 collider centre of mass energies 0.5, 1 and 3 TeV versus the Higgs boson mass in GeV.

To estimate the rate for e+e-→ZX, one needs the width of X into ZZ. This width can be at most equal to the measured experimental width, of order 45 GeV according to ATLAS, to be compared to the SM HZZ width of 70 GeV. From this argument and figure 3 from [20] one expects that the cross section e+e-→ZX at 1 TeV should be <0.5 fb.

From LHC measurements at 8 TeV, one has an upper limit on ΓZZ/Γγγ: **<12fb/1fb~10**. Even taking our largest assumption on the photon width, 45 MeV, the ZZ width would be ~0.5 GeV hence a cross section ≤0.01 fb, marginally measurable.

I therefore conclude that this production mode is unlikely to give the large amount of data needed to study X(750) in detail. The same is true if one uses WW fusion, even operating at 3 TeV.

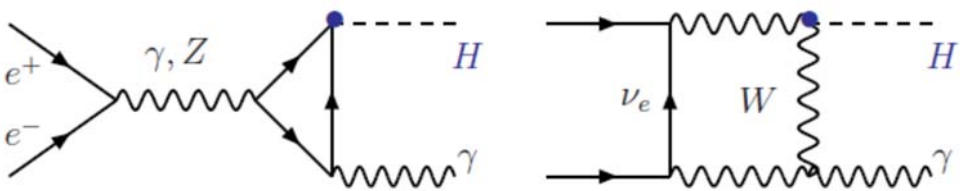

At first sight one would expect Xγ to be also negligible, as Hγ with respect to HZ. This is not the case since, recalling figure 2, one notices that Γγγ is at least 50 times larger than for a SM Higgs. From [21] one can deduce that at 1 TeV the cross section for H(750)γ is of order 0.0025 fb, meaning that it could reach ~0.1 fb for X(750)γ. This result agrees with [13] which evaluates this process using an effective field theory approach.

If X(750) turns out to be a wide resonance, figure 2 shows that Γγγ could be increased by up to 3 orders of magnitude and therefore reach 10 fb at 1 TeV, giving more than **10$^4$ events**.

In principle this reaction has a threshold near 750 GeV, however the threshold dependence is given by $(1-M_X^2/s)^3$, meaning that in practice one should operate at 1 TeV or more. At 1.5 TeV one gains a factor 5 on this cross section.

At 1 TeV, X is accompanied by a monochromatic photon with $E\gamma \sim 220$ GeV, which provides a very clean signature. This allows to measure the total cross section and therefore $\Gamma\gamma\gamma$. The $Z\gamma$ contribution can be separated from $\gamma\gamma$ by using an electron beam with longitudinal polarisation giving ALR.

Isolating any process, for instance ZZ, gives $\Gamma\gamma\gamma$BRzz hence BRzz. One can also determine BRinv, the invisible BR. It is however fair to say that unless this resonance turns out to be wide, the low cross section does not allow very high accuracies.

## II.2 Scenarios to produce e+e→X(750) at resonance

### II.2.1 Scalar case

Could one produce a (pseudo)scalar X in the s-channel at LC? This is possible, provided that the coupling of X(750) to quarks and leptons does not violate the chiral properties of the theory and can be done in two ways according to [22]. Either in the 'Higgs way' by having a coupling going like $(m_f/\Lambda)\bar{f}fH$ which leads to a negligible coupling to electrons, or by assuming that there is a pair of mass degenerate scalar and pseudo-scalar bosons, that I call H and A, with coupling h($\bar{f}fH$ +i$f\gamma_5\bar{f}A$ ), where h is an unconstrained constant. This set up allows to avoid undue chiral violations in loop contributions which come from these scalars. As an example, take the vertex correction to Z $\bar{f}f$ due to X exchange. It would show a perfect cancellation between H and A exchanges:

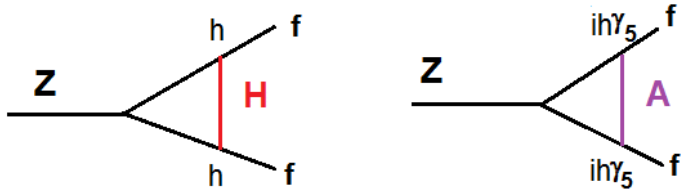

For the first diagram, the exchanged propagator is proportional to h²/(q²-m²), while the second has an opposite sign i²h²($\gamma_5$)² /(q²-m²)=-h²/(q²-m²), recalling that ($\gamma_5$)²=1.

The constant h should be small enough to accommodate the absence of an LHC signal in lepton pairs: $\Gamma(X \to l^+ +l^-) = \dfrac{2h^2 M_G}{8\pi}$, meaning that h<0.01. If these two particles were not perfectly degenerate in mass, one could explain an apparent large width of the signal. An LC collider operating at 750 GeV would be an ideal tool to study such a scenario.

### II.2.2 Tensor case

X(750) could alternatively be a heavy **graviton,** [23] to [27], which could in certain cases couple to e+e- [24]. This reference claims a similar width as for $\gamma\gamma$. This statement depends on the details of the Randall Sundrum (RS) model. Initially, when it was assumed that all SM particles live on the so called TeV brane, the KK graviton was expected to couple democratically to all fermions, hence to e+e-. In versions where SM particles propagate in the 'bulk', the KK graviton is preferentially coupled to top quarks and very weakly coupled to light fermions. Only future data from LHC can give us an answer.

Assuming a narrow resonance and a Gaussian beam spread, this process has a cross section:

$$\sigma = 2\pi^2(2J+1)\dfrac{\Gamma(X \to e^+e^-)}{\sqrt{2\pi}\sigma_M M_X^2}$$

With Γee=0.13 MeV, as in [24], and a beam spread of 1%, one expects $\sigma_X$ ~ 1 pb, to be compared to the point like cross section ~180 fb at 750 GeV. Initial state radiation, ISR, should reduce this value by a factor F=$(2.35\sigma_{Mx}/M_x)^b$ where b~0.13. For $\sigma_{Mx}/M_x$=1%, F=0.6.

An **X factory would then be possible** with all the advantages of LEP1/SLC, in particular a precise access to invisible modes.

A more complete study of this scenario will be given in section IV.

# III. A photon collider PLC

This possibility has been studied since a long time [28] together with the e+e- LC projects, having in mind a precise study of γγ→H(125). In this note, I do not intend to present the technicalities of this project, thoroughly covered since the TESLA TDR.

At present, it is felt that one can achieve a precise measurement of the SM Higgs coupling to two photons by using the measurement of the ratio H→γγ/H→ZZ given by LHC in combination with the precise determination of the ZZ coupling given by e+e- colliders through the reaction e+e-→HZ.

What can we expect from a **photon-photon collider**? One has [20] (where λi are the photon helicities which can be tuned to maximize the cross section):

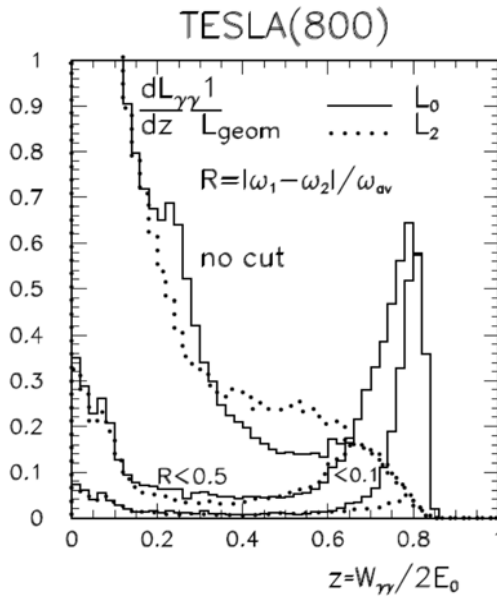

$$\sigma(W) = 8\pi(2J+1)\frac{\Gamma(X \to \gamma\gamma)\Gamma(X \to F)}{(W^2 - M_X^2) + M_X^2 \Gamma_X^2}(1+\lambda_1\lambda_2)$$

At resonance $\sigma_{tot}$=16πBR(2γ)/M²=36nbBR(2γ). If one assumes $\Gamma_X$/M=0.06 $\Gamma_{\gamma\gamma}$/M=6 10$^{-5}$ or BR(2γ)=10$^{-3}$, this gives 36pb. In the following we will show that such a solution provides a viable photon collider scheme.

It is quite clear from figure 4 that the beam spread can be narrowed by applying an **acollinearity cut**. The **energy reconstruction of the final state** could play a major role to eventually measure the width of this resonance in case the ATLAS indication is real. This spectrum is obtained for polarized photons such that Jz=0. For a graviton, it would be equally possible to obtain a similar spectrum for Jz=2.

*Figure 4: Differential predicted luminosity for a photon collider in terms of z, the centre of mass energy divided by the e+e- energy for two choices of the total helicity of the photon beams.*

One can compute the rate by convoluting the cross section distribution with this luminosity curve using the approximate formula [20] valid for a narrow resonance:

$$\sigma(W) = 4\pi^2(2J+1)\frac{\Gamma(X \to \gamma\gamma)BR(X \to F)}{M_X^2}(1+\lambda_1\lambda_2)\delta(W-M_X)$$

Note that photon polarisation can increase by a factor ~2 the production rate. For a scalar resonance, It also decreases the light fermion backgrounds as will become evident in III.1.

In figure 4, $L_{geom}$ corresponds to the geometrical luminosity as defined in [28], $L_{geom}\sim 2\,10^{35}\,cm^{-2}s^{-1}$ at ~1 TeV. This concept assumes that contrary to e+e-, round beams are possible since AT A plc beam-beam interaction is reduced.

On has $dL_{\gamma\gamma}=0.58 L_{geom}\,dW/2E_0$ where $2E_0\sim 1$ TeV. Using above formula for the cross section, the rate is therefore given by:

$$Rate = 4\pi^2(2J+1)0.58 L_{geom}\frac{\Gamma(X\to\gamma\gamma)}{2E_0 M_X^2}(1+\lambda_1\lambda_2)$$

**N$_X$/sec=6600(2J+1)($\Gamma\gamma\gamma$/2E$_0$)**

For $\Gamma\gamma\gamma/M=6\,10^{-5}$ one gets 0.3 evts/s or 1000evts/hour. However if one assumes that $\Gamma\gamma\gamma\sim$MeV then this rate drops to ~30 evts/hour, which, as said previously, constitutes a minimum for a scalar resonance. For a tensor, $\Gamma\gamma\gamma$ can be smaller but this is compensated by the (2J+1) factor.

## III.1 SM backgrounds at PLC and the radion

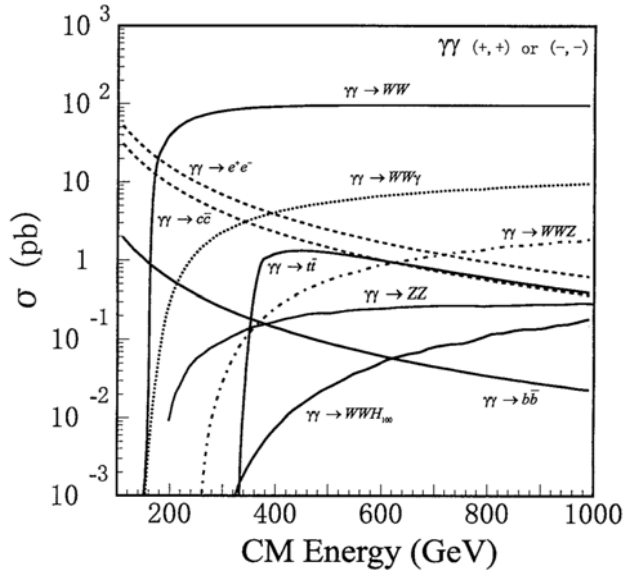

Figure 5: Photon-photon cross sections in pb versus the centre of mass energy in GeV assuming zero total initial helicity.

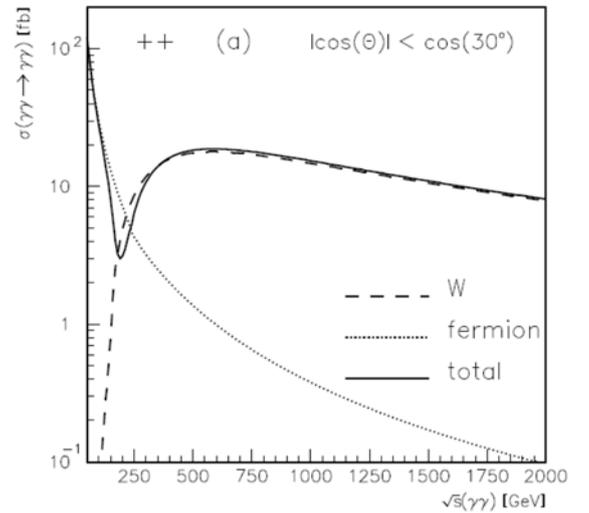

Figure 6: Cross section in fb for the process $\gamma\gamma\to\gamma\gamma$ versus the com energy in GeV. At 750 GeV the dominant contribution comes from the W loop.

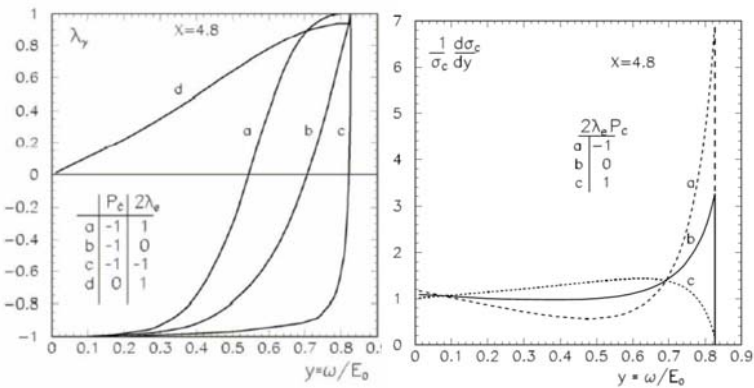

Figure 7: On the left, energy distribution and on the right average helicity of the backscattered photons versus their energy divided by the electron energy for various polarisations of the electrons and of the initial photons and electrons.

The SM backgrounds are shown in figure 5 for the scalar case.
Missing are the processes $\gamma\gamma\to\gamma\gamma$, $\gamma\gamma\to gg$, $\gamma\gamma\to Zh$ and $\gamma\gamma\to hh$ which are relevant to this study. The first one can be found in [29] and [30]. It is large, ~10fb, mainly due to the WW loop (as can be guessed from the very large cross section $\gamma\gamma\to WW$) as shown in figure 6.

On the contrary the process $\gamma\gamma\to gg$ has a cross section below 1fb, as deduced from [31]. The Zh process can be found in [32] and has a cross section of ~0.03 fb. The hh process is evaluated in [33]. The SM background is high for some relevant modes like WW and tt and requires a **detailed experimental** **study**. Here I will simply sketch the main features. If $\Gamma/M\sim 0.06$, the precise **detector particle energy flow resolution** should allow to measure the width of the resonance.

The fermion pair background is governed by the following formula from [34]:

$$\frac{d\sigma}{dx} = \frac{2\pi\alpha^2 e_f^4 N_c \beta}{s(1-\beta^2 x^2)^2}\left\{(1+\lambda_1\lambda_2)(1-\beta^4)+(1-\lambda_1\lambda_2)\beta^2(1-x^2)(2-\beta^2(1-x^2))\right\}$$

where $x=\cos\theta^*$. The photon beam distributions are shown in figure 7 from [28]. Spectrum **a,** which uses polarized electrons and polarized photons to generate the photon beam through Compton backscattering, has the best flat top for beam polarisation. In this study, I took **<$\lambda_1\lambda_2$>~0.9**.

For a scalar resonance, the second term is suppressed by beam polarisation while the first term is suppressed by $1-\beta^4$ for light fermions. For a **tensor resonance** with helicity $\pm 2$, the second term dominates, implying a larger background from light fermions.

This cross section is forward peaked and an angular cut gives an additional suppression.
The top quark background cannot be suppressed given that $\beta_t \sim 0.9$.

To derive the background rate one approximate the luminosity distribution shown in figure 4 by:
$dL/dz=0.58 L_{geom}\exp(-z^2/2 s_z^2)$ where $z=(W-W_R)/2E_0$, $L_{geom}=2 10^{35}$cm-2s-1. The beam effective spread, $s_z$, is reduced to ~2.5% by a collinear selection of the final states, as explained in [28].

One then gets:

$$Rate = 0.58 L_{geom}\sigma\sqrt{2\pi}s_z$$

where $\sigma$ is the SM background cross section taking into account beam polarisation effects in the scalar mode and appropriate angular selections.

For a scalar resonance, one assumes a narrow resonance with $\Gamma_{\gamma\gamma}=1$ MeV, that is a rate of ~30 events/hour. JJ means light quarks and gluon only, excluding bb which can be identified by b tagging.

| Process | JJ | bb | tt | ee/μμ/ττ | γγ | Zγ | ZZ | hh | WW | Zh |
|---|---|---|---|---|---|---|---|---|---|---|
| Back. rate/h | 1.2 | 0.04 | 20 | 1 | 0.5 | 0.5 | 0.5 | <0.01 | 200 | 0.03 |
| BR 5 sd % | 0.25 | 0.1 | 1 | 0.2 | 0.15 | 0.15 | 0.15 | 0.02 | 3 | 0.01 |

The second row gives the estimated background per hour, while the third row gives the signal BR needed to measure a five standard excess with respect to the background assuming a full year running time (24*365 hours).

Most LHC present limits are irrelevant with the exception of the ee+μμ. In the future the LHC limits will go down but there are clearly several modes with outstanding sensitivity for PLC like tt, bb, hh and gg. For jet-jet modes, one will be unable to distinguish gluon jets from light quark jets but most models predict that the latter is negligible.

| | $\Gamma_t$ GeV | BRγγ % | BRgg % | BRzz % | BRww % | BRtt % | BRhh % |
|---|---|---|---|---|---|---|---|
| 1 | 0.201 | 0.5 | 89 | 2.6 | 2 | 1.7 | 3.9 |
| 2 | 0.151 | 0.7 | 96 | 1.0 | 1.4 | 0.27 | 0.6 |
| 3 | 0.106 | 0.9 | 94 | 1.4 | 2 | 0.38 | 0.9 |

Reference [5] provides various benchmarks for the **radion scenario.** They predict rates consistent with the sensitivity of PLC, with the exception of the bb and Zγ modes which have very low BR. As an example, first three **benchmark scenarios** taken from [5] are shown in above table, with all modes providing an adequate sensitivity, with the exception of WW which will be barely measurable.

### III.2 Polarized photons and spin parity: Scalar or pseudo-scalar?

We just saw how beam polarisation can enhance the visibility of a scalar resonance by doubling its cross section and decreasing by an order of magnitude the light fermion backgrounds.

This polarisation can also help in identifying this object which can be a scalar, a pseudo-scalar or both. For the Higgs boson itself, this is done by using the ZZ* mode from an angular analysis involving the decay products but, so far, one has not observed this mode for X(750). The mode into two photons in principle also allows for such a discrimination by measuring the photon polarisation by either using virtual or converted photons, but this requires cumulating huge statistics.

As already studied for the reaction γγ→H/A in [35], one can show that transversally polarized initial photons will allow to discriminate between CP=1 and CP=-1 final states. One can construct a simple asymmetry by taking the two transverse polarisations either perpendicular ⊥ or parallel //. If they are // one will only produce H while only A is produced for perpendicular polarisations.

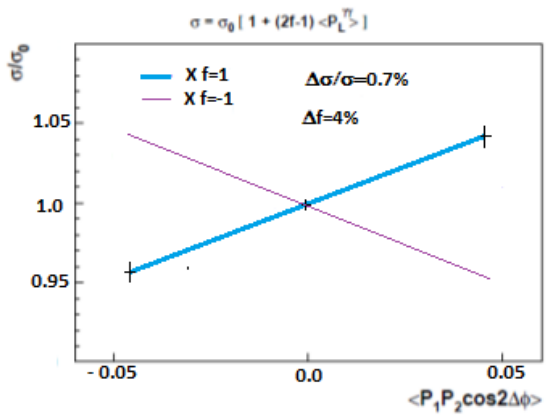

*Figure 8: Expected dependence of the cross sections on transverse polarisation for a scalar X (f=1) or a pseudoscalar X(f=-1).*

Figure 8 shows the expected dependence of the cross section assuming that 20% of the data have been taken with transversally polarized photons with a conservative transverse polarisations of 20% per beam, as deduced from [28].

If gg mode is the dominant decay mode, this channel will be contaminated by light quark final states. Without longitudinal polarization one simply has:

$$\frac{d\sigma}{dx} = \frac{2\pi\alpha^2 e_f^4 N_c \beta}{s(1-\beta^2 x^2)^2}\left\{(1-\beta^4)+(1-x^2)(1+x^2)\right\}$$

as deduced from previous formula. Cutting at |x|<0.8, this background gives 10 event/h to be compared to 30/h for the signal which, in a scenario where gg is the dominant mode, corresponds to an excellent s/b situation. The **fraction of CP even** scalar f is measured with a **4% error.**

The 20% transverse polarisation assumes a standard **laser wavelength** of a μm as described in [28]. It is also conceivable to use a laser with ~**2μm**, in which case one could achieve a polarisation of 35%, hence an accuracy on **f at the % level.**

# IV The tensor scenario

In this section one assumes that X(750) is a tensor, J=2 resonance, for instance a heavy KK graviton as in RS. RS usually predicts a tensor particle which couples to gluon-gluon but with a mass above 3 TeV.

However several references assume that this particle could be a KK graviton as in [23] to [27].

RS usual mass hierarchies can indeed be modified by introducing '**brane kinetic terms**'. It is possible to have a light KK graviton while keeping the next KK excitation out of reach of LHC. Also the vector KK particles, if not influenced by these kinetic terms, could be much heavier, allowing to satisfy EW constraints. Note however that there are escapes to these EW constraints, as explained in section IX.

X(750) can then couple to γγ and one can go back to section III for the expected luminosity and backgrounds. The main difference comes from the beam polarisation which suppresses less efficiently the light fermion background.

The angular distribution of two processes, γγ→ff and γγ→γγ/gg, allows to unmistakably identify the spin of this particle as explained in IV.1.

In some models [25], a tensor X can also couple to e+e-, a statement which obviously waits for confirmation from LHC

## IV.1 Angular distributions

### IV.1.1 gg↔γγ

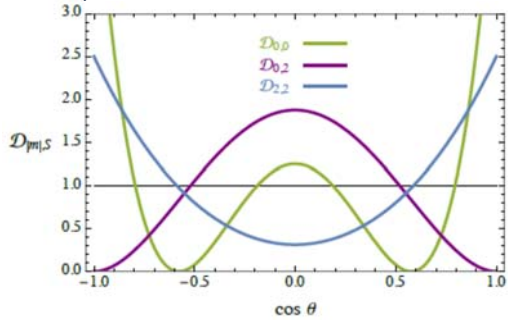

*Figure 9: Angular distributions for γγ↔gg for a tensor resonance.*

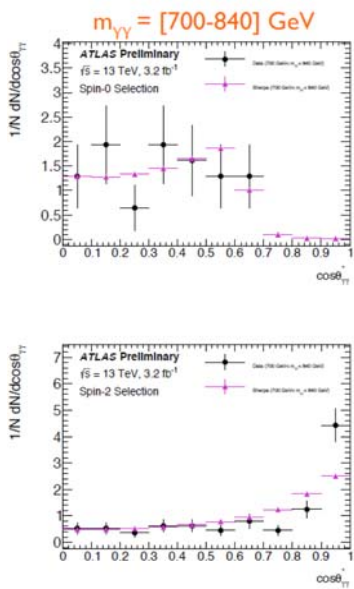

Reference [36] provides the angular distributions expected for the process γγ↔gg which are both valid for LHC and PLC. Of key importance for LHC is this angular distribution gg→γγ since it allows LHC to identify a tensor resonance. For the KK graviton case only ±2 helicities are relevant, meaning $\mathcal{D}_{2,2}$.

*Figure 10: cosθ* distribution of the decay of X(750) into two photons as seen by ATLAS for the two analyses presented in Moriond EW 2016. On top one sees the standard analysis for a scalar decay and below the new analysis accommodating a tensor.*

Figure 10, borrowed from the recent analysis presented by ATLAS at Moriond EW [1], illustrates this point. This figure shows that for the spin-zero analysis, there are no events for |cosθ*|>0.7 while for the new analysis one sees that the data and the background tend to peak at large |cosθ*|. For the latter this is expected since the dominant contribution comes from quark anti-quark annihilation. Remarkably, if LHC observes a clear sign of forward peaking, this would not only reject the scalar hypothesis, but also prove that the initial sate is necessary gg and not quark anti-quark.

This can be seen in figure 11 and is simply interpreted as Jz mismatch between initial state Jz=±1 for light fermions and the final state Jz=±2 for gluons and photons.

Within RS, one predicts mainly gg→ hh, $W_L W_L$ and $Z_L Z_L$. They should correspond to $\mathcal{D}_{2,0}$.

## IV.1.2 qq/ee↔γγ/gg

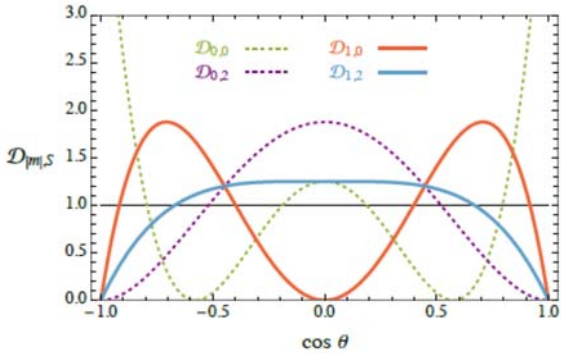
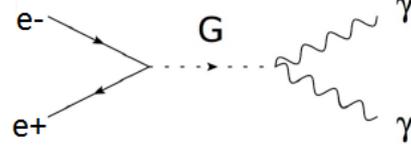

A graviton could also couple directly to e+e- . [36] gives the angular distributions for qq→γγ also valid for ee→γγ. Assuming helicities h=±2 at the photon vertex, only $\mathcal{D}_{1,2}$ is relevant as shown in [36]. If X(750) originates from quark annihilation, there is no forward peaking and the spin 0 analysis from ATLAS is adequate.

Figure 11: Expected angular distribution for e+e-->X(750)->e+e- for a tensor particle. In practice only 1,0 and 1,2 distributions are relevant, while others are helicity suppressed

## IV.1.3 ee↔ff

Figure 12: Expected angular distribution for e+e-->µ+µ- for a graviton resonance.

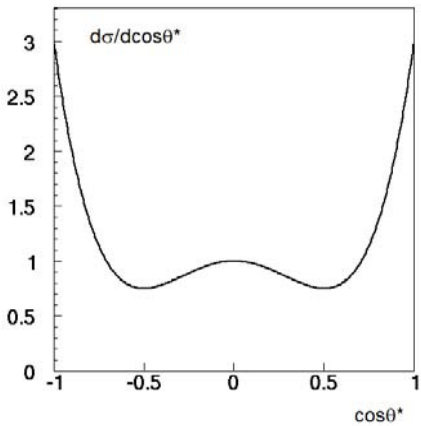

Using the formalism provided by [36] one can easily derive that the corresponding angular distribution varies like:

$\mathcal{D}_{1,1} \sim 4z^4 - 3z^2 + 1$ where $z = \cos\theta^*$

in agreement with [37].

The shape of this distribution, shown in figure 12, provides a strong signature for a tensor particle and illustrates again the need for a good forward coverage. For top quarks, one should use the distribution:

$4z'^4 - 3z'^2 + 1 - (1-\beta_t^2)(1-4z'^2)$ where $z' = \beta_t \cos\theta^*$

## IV.3 Expected rates with a PLC

In [25] several RS scenarios are proposed which are summarized in the following table. Scenario 'IR' offers the possibility to produce X(750) in a e+e- and will be discussed as such in the next section.

|      | BRγγ % | BRjj % | BRzz % | BRww % | Rzγ % | BRhh % | BRtt % | BRbb % | BRee % | Γγγ MeV | Γtot MeV | Evt/h | Rate/year 10[5] |
|------|--------|--------|--------|--------|-------|--------|--------|--------|--------|---------|----------|-------|-----------------|
| IR   | 4.3    | 66     | 4.8    | 9.5    | 0     | 0.3    | 5.1    | 6.4    | 2.1    | 0.28    | 0.28     | 28    | 3               |
| MIN  | 8.5    | 68     | 7.9    | 16     | 0     | 0      | 0      | 0      | 0      | 0.14    | 1.7      | 17    | 1.5             |
| MED  | 7.0    | 61     | 7.8    | 15     | 0     | 0.4    | 8.3    | 5.2    | 0      | 0.17    | 2.4      | 20    | 1.8             |
| MAX  | 0.5    | 4.5    | 2.9    | 5.6    | 0     | 1.4    | 85     | 0.4    | 0      | 2.5     | 500      | 290   | 26              |
| GMAX | 2.3    | 0.5    | 12     | 21     | 1.1   | 6.9    | 56     | 0.04   | 0      | 25      | 1130     | 2900  | 260             |

Backgrounds are estimated applying the methods already described with the following results:

| Process    | jj | bb  | tt  | ττ | µµ | γγ  | Zγ | ZZ | hh | WW  |
|------------|----|-----|-----|----|----|-----|----|----|----|-----|
| back/hour  | 20 | 0.3 | 100 | 34 | 3  | 0.5 | 5  | 12 | 1  | 700 |

Recalling the formula:

$$\frac{d\sigma}{dx} = \frac{2\pi\alpha^2 e_f^4 N_c \beta}{s(1-\beta^2 x^2)^2}\left\{(1+\lambda_1\lambda_2)(1-\beta^4) + (1-\lambda_1\lambda_2)\beta^2(1-x^2)(2-\beta^2(1-x^2))\right\}$$

for the tensor case, where $\lambda_1\lambda_2\sim-0.9$, one expects a larger background than for the scalar case, from the predominance of the second term in {}. A forward cut, $|\cos\theta|<0.8$, is applied reducing the geometrical efficiency to 59% for the gg and $\gamma\gamma$ channels which are forward peaked (curve $\mathcal{D}_{2,0}$).

From this table it is straightforward to deduce that each of the scenarios of [25] can be fully tested, meaning that the predicted modes can be observed in all cases with more than five standard deviations significance. In certain cases LHC will be able to observe new modes but when the BR are at a few per cent level this will not be possible as for tt,bb,jj and hh.

## IV.4 Expected rates with an e+e- collider

In the **scenario IR** described in reference [25] where all SM particles belong to the TeV brane, the graviton couples to e+e- with half the width into two photons. One can simply predict the expected widths and BR of the SM in the following way.

| $\Gamma$ee MeV | $\Gamma_t$ MeV | Bree % | BR$\gamma\gamma$ % | BRzz % | BRww % | BRz$\gamma$ % | BRhh % | BRtt | BRbb | BRjj |
|---|---|---|---|---|---|---|---|---|---|---|
| 0.14 | 5.7 | 2.1 | 4.3 | 4.8 | 9.5 | 0 | 0.3 | 5.1 | 6.5 | 66 |

For a narrow resonance and unpolarised beams, the total cross section in e+e- is:

$$\sigma(W) = 2\pi^2(2J+1)\frac{\Gamma(X\to e^+e^-)}{M_X^2}\delta(W-M_X)$$

Assuming a narrow resonance and a Gaussian beam mass distribution with $\sigma_M$:

$$\sigma = 2\pi^2(2J+1)\frac{\Gamma(X\to e^+e^-)}{\sqrt{2\pi}\sigma_M M_X^2}$$

With a beam spread of order 1% and $\Gamma_{ee}$=0.14MeV, this gives 1 pb. Radiative corrections further reduces this cross section by a factor ~2. The typical integrated luminosity is or order 2ab$^{-1}$, meaning **10$^6$ collected events.** Backgrounds are given below noting that the SM production of hh is found in [38]. As explained in the next section it will be possible to measure $\Gamma_{ee}$ and all other widths.

| Type | $\mu\mu$ | JJ | bb | tt | $\gamma\gamma$ c<0.8 | WW (c<0.8) | ZZ (c<0.8) | hh |
|---|---|---|---|---|---|---|---|---|
| SM fb | 200 | 1000 | 320 | 300 | 250 | 400 | 50 | <0.02 |
| X fb | 10 | 300 | 30 | 20 | 20 | 40 | 20 | 1 |
| dBR/BR % | 8 | 0.3 | 2 | 2 | 2 | 1 | 1 | 2 |

Forward cuts on WW, ZZ and $\gamma\gamma$ can be applied without any significant losses. These cuts are irrelevant for light fermions since there is no t-channel exchange.

## VI.4 Total width and invisible width

If the total width can be measured, the situation is as good as LEP1: all partial widths can be extracted, including the invisible width by comparing the total width to the sum of partial widths.

This situation corresponds to the purple diamond in figure 2. The rate could then be of 3000 evts/hour. A major question would be to understand the origin of this large width: new exotic particles, visible or invisible?

In the models presented so far, the total width is too narrow to be directly measured. If X(750) can be produced in e+e-, the situation will be analogous to the J/$\psi$ at SPEAR where one could deduce simply the invisible width by using the standard radiative return method by running above the resonance. This method provides $\Gamma_{ee}BR_{inv}$ at the % level (or an upper limit on this quantity). The data taken on the resonance should give $\Gamma_{ee}BR_{vis}$ by summing all final states with a similar accuracy. The sum of these two terms give $\Gamma_{ee}$. The next step is to measure $\Gamma_{ee}BR_{\mu\mu}$ which gives $BR_{\mu\mu}$ with an accuracy of ~10% (see previous section). Assuming universality, $\Gamma_{ee}=\Gamma_{\mu\mu}$, one deduces the total width which is the denominator of $BR_{\mu\mu}$, with an accuracy of 10%.

# V (More) Speculations

Here, I wish to make further comments on two privileged interpretations which give a wide scope like WTC (walking TC) and RS (Randall Sundrum).

## V.1 Walking technicolor

In WTC (see for instance [8]) X is a **techniscalar** $P^0$.

| Nc | 3 | 4 |
|---|---|---|
| $\Gamma_{tot}$ GeV | 1.2 | 2.1 |
| BR($P^0$ -> gg) % | 99.8 | 99.8 |
| BR($P^0$ -> $\gamma\gamma$) % | 9.7 x $10^{-2}$ | 9.7 x $10^{-2}$ |
| BR($P^0$ -> $Z\gamma$) % | 5.3 x $10^{-2}$ | 5.3 x $10^{-2}$ |
| BR($P^0$ -> ZZ) % | 9.7 x $10^{-2}$ | 7.3 x $10^{-2}$ |

One has $\Gamma_{\gamma\gamma}$=2 MeV which gives ~60 events/hour at a photon collider. Recalling section III.1, one concludes that ZZ and $Z\gamma$ are detectable.

In reference [17] it is argued that a technipion can have an enhanced coupling to $\gamma\gamma$ due to anomalous couplings. Therefore a further enhancement of $\Gamma_{\gamma\gamma}$ is possible which could improve the situation.

## V.2 X(750) ➔ hh

As noted in the introduction, this channel is of particular importance for extended scalar sectors proposed to explain baryogenesis [11]. A natural consequence is that gg➔X➔2h could have a large cross section, above 20 fb, very near the present limit of exclusion reached at Run 1 and certainly within reach at Run 2. The SM process are:

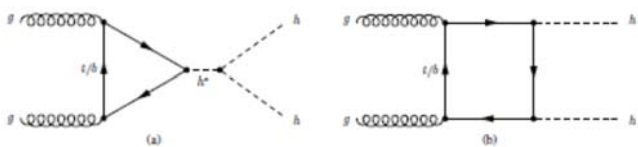

The total SM cross section ~10 fb is almost negligible at 750 GeV. Figure 13 from [39] indicates a slight excess nearby, due to the bb$\gamma\gamma$ channel.

In RS (see for instance [5]) a **radion** has $\Gamma_{\gamma\gamma}$~1 MeV. The decays depend on the Ricci-Higgs mixing $\xi$ parameter which needs to be close to the conformal limit 1/6 to be consistent with LHC results. The gg decays dominates followed by hh which can reach a BR at the few % level, therefore measurable and allowing to distinguish between RS and WTC. This is also true for a heavy **graviton** [25].

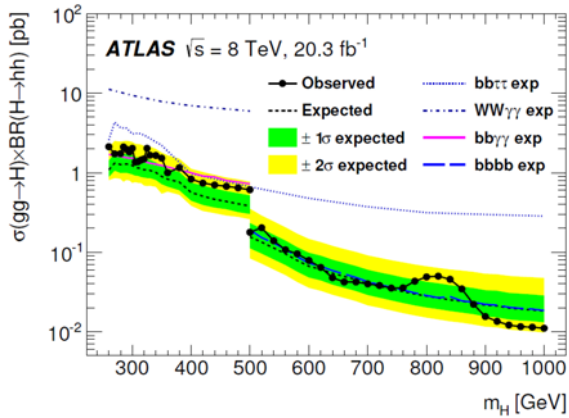

*Figure 13: Expected and observed limits from ATLAS for a heavy resonance decaying in a pair of h(125) bosons.*

## V.3 Complementary signals in RS

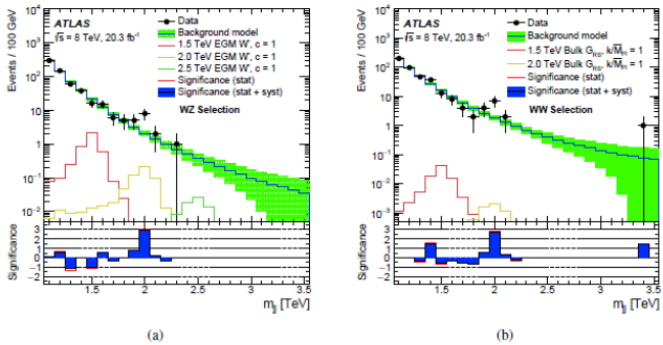

ATLAS and CMS, at 8 TeV, have also observed a few 2-3σ bumps in WW/ZW/ZZ/Wh, around 2 TeV, as recalled by figure 14. These bumps are not confirmed at 13 TeV but this is inconclusive given the low luminosity recorded. Reference [40] provides an interpretation of the bumps in terms of RS.

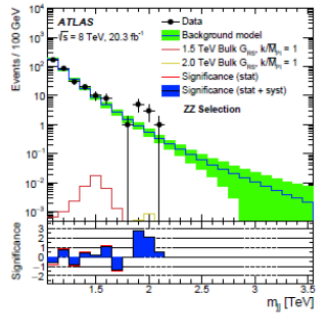

Sum of all modes

17 events 9 back

*Figure 14: The 3 plots recall the observations made by ATLAS on the search for heavy resonances decaying into WW, ZZ and ZW [41].*

➔ **Such signals, if confirmed, would call for an e+e- collider with 2 TeV center of mass energy**

# VI Higgs and top properties

The two models under consideration, RS and WTC, predict measurable alterations of the Higgs properties which should be observed in e+e- colliders.

## VI.1 Walking TC

For WTC, the Higgs boson is interpreted as a 'technidilaton'.

| Mode | gg | bb | WW | ZZ | ττ | γγ | Zγ | μμ |
|------|----|----|----|----|----|----|----|----|
| BR % | 75 | 20 | 3.6 | 0.40 | 1.20 | 0.07 | 0.005 | 0.004 |

Given the recent combination of ATLAS and CMS, there seems to be incompatibility with the data, in particular for what concerns γγ/ZZ and bb/ZZ as shown in figure 16.

No doubt that WTC can still improve to explain this discrepancy but, at the moment, it does not qualify.

## VI.2 Higgs and top electroweak couplings in RS

While the presence of a light radion or a KK graviton does not influence directly the Higgs sector, it suggests that vector-like fermions and recurrences from Z/W/g/γ could be light and therefore influence substantially this sector. Higgs couplings can indeed show very strong deviations if, as suggested by preliminary indications, the 1st KK recurrences of Z and W are at 2 TeV. This is shown in figure 15 from [42] which also shows that the expected deviations very much depend on unknown Yukawa couplings of the new fermions. Even if, as suggested by [25], the vector particles are much heavier, **beyond LHC reach**, figure 15 shows that deviations are measurable at a LC, which again shows the power of such measurements when performed at a LC.

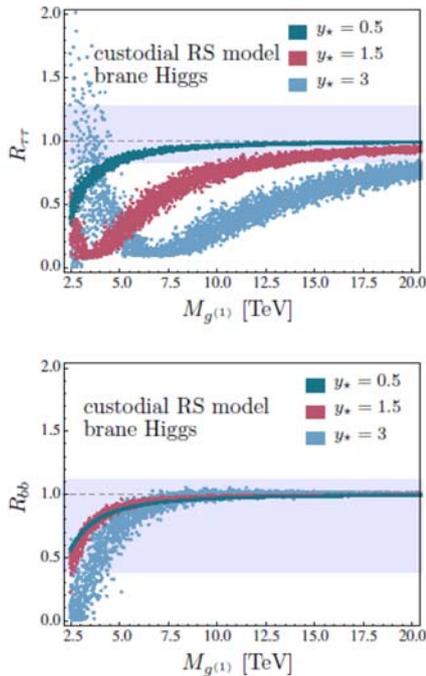

*Figure 15: Examples of RS predictions for the expected rate of ττ and bb versus the mass of the first KK gluon excitation. y\* is a parameter related to Yukawa couplings.*

LHC measurements do not allow to separate the various widths but are sensitive to combinations of the type Rxx=ΓggΓxx/Γtot (forgetting the fusion contribution) with each term varying rapidly when the KK gluon mass gets below 5 TeV. If the signal observed at 2 TeV is real, then one finds that the individual width may strongly deviate from the SM while, through compensations, the ratio R remains close to 1. For the bb mode however, with Γbb/Γtot~1, Rbb depends mainly on Γgg. Ratios of Rxx only depend on ratios of partial widths and may reveal earlier some hidden trends.

Present observations are given in figure 16 [8]. From the combination of ATLAS+CMS data, the largest deviation is seen on the bb/ZZ. This type of effect naturally occurs within RS.

Assuming that the first KK sequence of W/Z/g is at 2 TeV, and adjusting properly the Yukawa parameters to be in agreement with LHC data, one obtains the corrections factors shown in figure 17.

Note that for computing the correction factors for γγ and gg, there is a sign ambiguity in RS. While the so-called 'Brane' model fails to reproduce γγ/ZZ, it agrees with the 'Bulk' model.

Figure 17 provides the correction factors to the SM widths such that Γi=Ci² Γism. In e+e- colliders, it will be possible to measure separately these widths, including the total one, with an accuracy at a few % level, which should lead to large and interpretable deviations. Comparisons of LHC and LC accuracies can be found in [43]. These accuracies are still a moving target, but, for sure, come out one order of magnitude better at LC that those foreseeable at LHC. As noted in [5], radion-Higgs mixing will at most affect by a few per cent the Higgs BR for gg induced processes. Even with e+e- accuracies, this effect seems challenging to distinguish from the dominant RS contributions.

While these measurements alone cannot provide unambiguous answers, it is clear that direct discovery of massive resonances by LHC will tremendously help in getting an interpretation of these deviations.

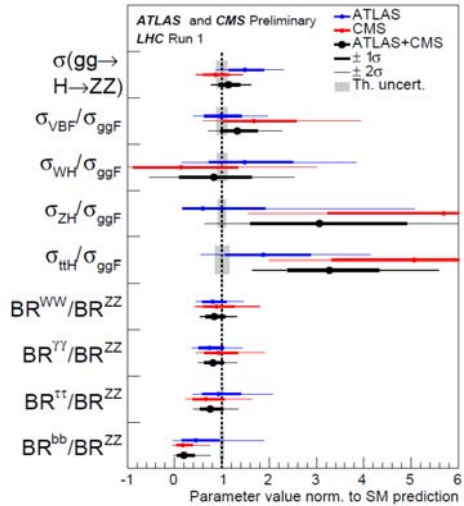
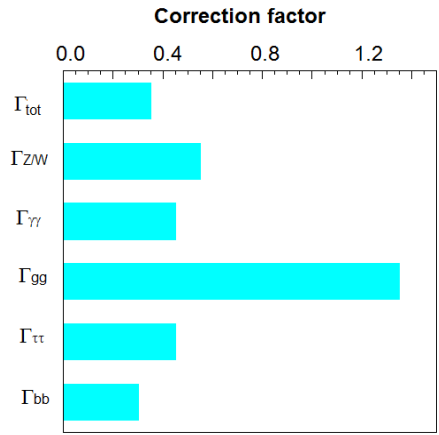

Figure 16: Various observables of cross sections and ratios of BR for the SM Higgs boson from ATLAS and CMS and combination of the two experiments.

Figure 17: Correction factors Ci to be applied to the various Higgs modes where $\Gamma_i = C_i^2 \Gamma_{iSM}$.

For top physics, the RS scenario has also been discussed and illustrated. In [44] it was shown that an e+e- machine with polarized beams can provide % accuracies on electroweak couplings, well beyond LHC. Reference [45] provides examples of BSM theories, among which RS, which can be thoroughly studied by these measurements.

## VI.3 ttH and ttW anomalies

Both ATLAS and CMS observe modest but similar excess in ttH [8] and ttW [46], [47]. gluon$_{KK}$→t't' or t't with t'→tH, gives ~ the right amount of events for $m_{t'}$~1 TeV. The analysis for ttW uses like sign lepton signature which naturally occur in b'b' final states as shown by the diagram below.

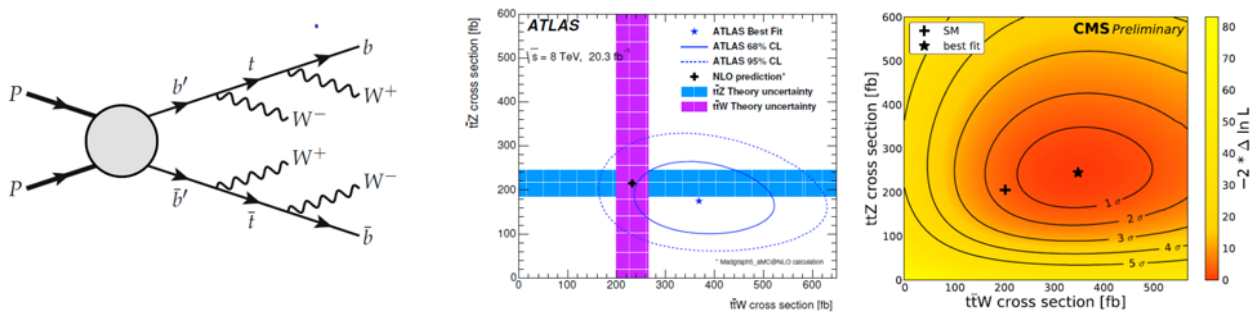

Figure 18: ttZ and ttW results for ATLAS and CMS at 8 TeV.

### VI.4 Higgs self-coupling

As mentioned in the introduction, usual extensions of the Higgs sector motivated by baryogenesis tend to affect the Higgs SM potential and therefore the Higgs self-coupling. Unfortunately this measurement will remain limited for a long time even assuming deviations reaching ~50%. Reference [11] gives an example of such a scenario.

Common wisdom tells us that it will require a TeV e+e- collider or a 100 TeV pp collider to reach the 10% precision needed for a meaningful result.

## VII X(750) and our universe

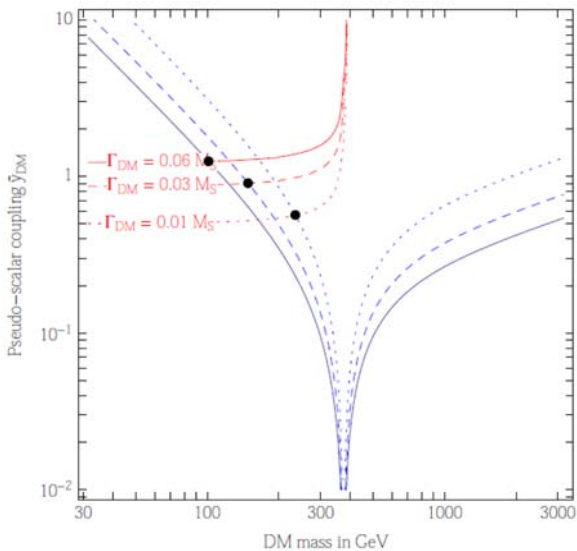

Figure 19: Allowed regions for a DM scenario involving X treated as a pseudocalar resonance.

This new particle is likely to be 'useful' in some ways. As already mentioned, it could allow a first order electroweak transition at the beginning of our universe and provide the needed mechanism for baryogenesis[11].

It could also play a major role in providing a 'portal' for dark matter annihilation in our universe. This possibility has attracted large attention and is presented in reference [15], from which I borrow figure 19 which assumes that X(750) is a **pseudoscalar** particle, avoiding any constraint from direct detection. Indeed the coupling to matter then becomes spin dependent and has a suppressed cross section. Interestingly the 3 allowed solutions shown by points in figure 19 give, for 3 values of the DM width $\Gamma_{DM}$, the right DM content of our universe.

Given that X (750) could have a substantial BR in $\gamma\gamma$, one can speculate [48] that indirect detection could observe a monochromatic peak through the reaction DM+DM➔X(750)➔$\gamma\gamma$. In a recent past, a very uncertain signal from Fermi-LAT has been reported peaking at 130 GeV (see for instance [49]).

At colliders, one can hope to observe the presence of this mechanism given that X(750) would have a large invisible branching ratio.

The heavy graviton scenario allows similar possibilities which are described in [50]. If coupled to e+e-, X(750) invisible decays can be precisely measured as described in IV.4.

## VIII Predicted new particles

Most of these models, which interpret X(750), have in common that they predict new particles observable at LHC. Here is an incomplete summary.

- RS predicts new vectors, KK recurrences of g/Z/W/$\gamma$ but also extra symmetries with Z'/W' states with roughly the same masses. As stated in [40] these particles may have already been observed at LHC. If so, it would require a LC reaching 2 TeV to observe them in e+e-.
- If the 1st KK recurrence is indeed at 2 TeV, RS predicts HO KK vector recurrences with masses at ~4.5 TeV, i.e. within the reach of LHC.
- RS predicts a tensor state, KK recurrence of the graviton, which couples to two photons. This state is about 50% heavier than the Z' state and therefore beyond the reach of a PLC. This prediction is not unique since several references identify X(750) to a heavy graviton.
- RS not only predicts also fermionic KK recurrences of SM particles but also lighter fermions, with quantum numbers which differ from SM particles like T5/3, a coloured quark carrying a charge 5/3. From present LHC limits, one may infer that these particles are heavier than 500-700 GeV, therefore not necessarily beyond the reach of a future LC.
- WTC also naturally predicts a rich spectrum of new vector and axial bosons and has charged scalar bosons. Extra scalar bosons, charged and neutral, are also predicted within extended Higgs sectors.
- If X(750) is a single wide resonance and assuming that it couples through loops with weakly interacting fermions, one expects a large number of new vector-like fermions which could be light. To explain why they have been missed so far by LHC, one would expect that they do not carry colour as in [18] and [19]. These would of course be ideal preys for a LC.

# IX. Indirect constraints from precision measurements

These constraints constitutes for most BSM schemes a severe challenge more or less easily overcome.

TC has been considered 'dead' on several occasions on the basis of the Peskin-Takeuchi variable T and S. Revivals usually invoke several tricks which, for instance use a cancellation between the vector and axial-vector contributions to the S variable, by analogy with QCD which contains both a $\rho$ and an A1.

RS has to cope with about the same kind of challenges and usually invokes 'custodial symmetries' to do so, with the appearance of new Z'/W' states with no corresponding zero mode.

These indirect constraints are subject to uncertainties from higher dimensional non-renormalisable operators originating from the ultra-violet completion of the model. The latter can potentially lead to large (and hopefully compensating) effects; see for instance, the recent analysis of [51] in a similar context.

Reference [52] proposes a mechanism to suppress large contributions to S/T from warped models. Reference [53] specifically evaluates S/T contributions due to the radion (assuming that SM particles are on the TeV brane), and comes out with small numbers as long as the radion is heavier than the Higgs boson and has small mixing.

Reference [54] provides a scenario which allows a KK vector boson as light as 1.5 TeV for warped models.

From this brief survey of the RS world, one deduces that there is a **variety of approaches** to produce models with light KK particles without violating **precision constraints**.

In [42] one introduces protection symmetries – the so-called **custodial symmetries** – which imply the appearance of new types of bosons and fermions. As already mentioned, [52] reaches a similar result by modifying the **geometry** of the extra dimension.

To be compatible with a light KK graviton, as in [25], one introduces the so-called **brane kinetic terms**. **Brane curvature** [55] also offers similar possibilities.

Also considered is the possibility of a **bulk Higgs** scenario as in [56] which allows a natural suppression of the radion couplings to heavy vector bosons but maintains a substantial coupling to top pairs.

From this type of considerations, one gets the feeling that extra warped dimension models offer a beautiful playground for imaginative theorists. To achieve a convincing distinction between these scenarios (and many others!), **precision measurements** from LC or PLC, both on Higgs and X(750) decay modes, will be the indispensable complement to LHC heavy resonance discoveries.

If X(750) is a wide resonance, one usually requires a large set of new states in the loops and these states could contribute to indirect measurements. This feature has been recently discussed in [57] which predicts observable deviations at LHC13 in Drell Yan measurements.

## X. Some issues

A wide resonance? As noted for instance by [15], if X has $\Gamma/M \sim 0.06$ as indicated by ATLAS, this means quite a change.
One needs a much larger photonic width to maintain a reasonable BR which is awkward in usual models and one has to identify the final states contributing to this width. Given the LHC present limits, the top and bottom quark final states offer the best opportunity but one cannot exclude the presence of new particles visible or invisible. However, as noted in reference [58], to pass the monojet search for invisible decays, this contribution cannot explain the largest part of this width.

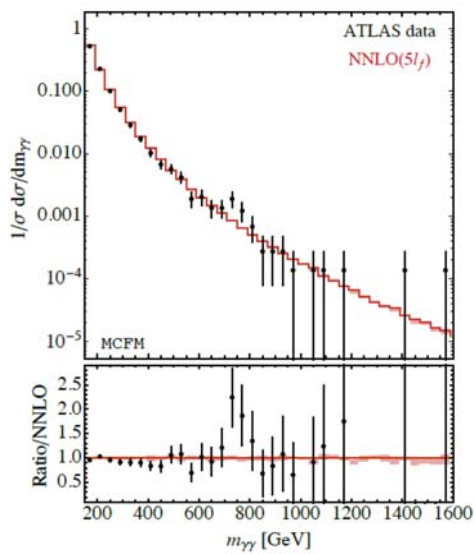

*Figure 18: Predicted $m_{\gamma\gamma}$ distribution at NNLO compared to ATLAS data.*

If this resonance is indeed wide, one should worry about possible mass distribution distortions due to interferences with the SM background. At LHC this will occur mainly in gg→tt where this background is huge, while for ZZ this effect should be smaller since the main contribution comes from light quark annihilation, therefore not interfering.

For the $\gamma\gamma$ mode at LHC, the SM background is large, s/b~1, but comes mainly from quark-antiquark annihilation which does not interfere with the putative signal, while the gg box diagram contribution is at the 10% level in the region of high masses [59].

For PLC and LC there should be interferences for all modes. It is premature to enter into these details at the present stage.

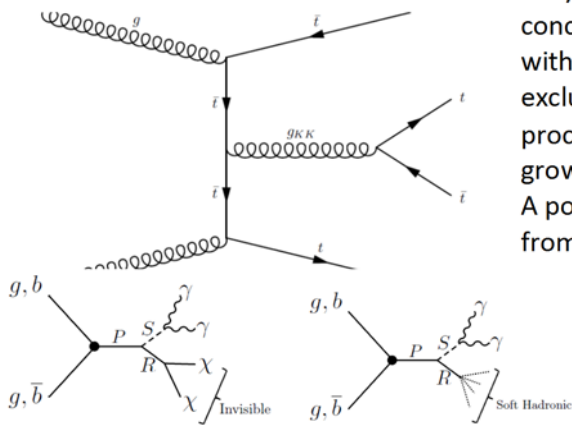

Why is there (almost) no signal at 8 TeV? From [15] one concludes that, within 2σ, the cross section is compatible with a growth of a factor ~5 between 8 TeV and 13 TeV. This excludes, for instance, that the main production mechanism proceeds through γγ annihilation since the corresponding growth is only ~2.

A possible answer could be that this object does not come from gg fusion but, e.g., from tt fusion in which case the gain with energy (replace $g_{KK}$ by X).

X could be pair-produced with an invisible heavy Y, meaning that the effective mass is much higher hence the large gain with energy [15].

These explanations seem incompatible with the data, as stated by ATLAS, since they would imply a measurable Ptmiss unbalance and/or extra jet activity, not observed in their data.

# Preliminary conclusions and prospects

- Some elements favour the hypothesis that this resonance comes from **gg→X→2γ**.
- If coupled to ZZ, this resonance could be produced in ZX at a TeV e+e- collider. Present limits suggest however that the cross section should be **below 0.1 fb.**
- An alternate would be to use the process e+e→Xγ which, given the expected value of Γγγ, could give a cross section of about **0.1fb** at 1 TeV and five times larger at 1.5 TeV. For a wide resonance which requires a dramatic increase of Γγγ, this cross section could be increased by two orders of magnitude.
- This discovery calls for a **photon collider** PLC which can in principle fit into a 1 TeV e+e- collider.
- Present constraints suggest that the cross section at resonance could be substantial, allowing giving between **30 to 3000 evts/hour**.
- Detailed studies are required to assess the observability of tt/WW/ZZ/hh/…which is a prerequisite to defend the need for such a machine. It appears however that using polarized photon beams and angular selections one can work out an excellent separation for most channels.
- With these accuracy a PLC can uniquely discriminate the various RS interpretation of X(750) in terms of a **KK graviton** [25] or a **radion** [5] by detecting the rare modes.
- This also applies to the **WTC** scenario.
- Observing tt/bb/jj or invisible modes at LHC looks very challenging and, for the models presented here, almost hopeless.
- A PLC with transversally polarized photons provides a unique opportunity to measure at a few % accuracy the **CP composition** of this new resonance.
- It is not excluded that a standard **e+e- LC** operating at **750 GeV** can do the job, in particular in the heavy **graviton scenario** where ~$10^6$ events could be produced. All measurable partial widths including the invisible width, could be measured with unsurpassable accuracy.
- If LHC provides the masses of the new particles, the deviations on **Higgs couplings** precisely given by an LC operating up to 500 GeV, will be easier to interpret, allowing a deep understanding of the BSM theory.

- There are weaker indications of vector resonances decaying into VV or Vh and vector-like fermions which seem to require an e+e- collider reaching **2 TeV**.
- In any case one still needs to understand much better what has been seen to assess the importance of an LC, eventually operated as a photon collider or, perhaps, as a standard e+e- collider. Given the cost and the technical challenges involved for a PLC, one needs to produce extremely strong physics arguments to go into that direction.

LHC should provide in 2016 the final word on the existence of this resonance. If confirmed, it will open a new era in particle physics and presumably well beyond, raising the question of the role of LC in this new domain.

In summary, the next questions will be:

- Wide or narrow?
- Spin 0 or 2?
- CP even, odd or both ?
- Which other modes?  hh, ZZ/WW, Z$\gamma$, tt,bb, ee/$\mu\mu$
- Other signals? VV and Vh states, vector-like fermions predicted in many schemes, heavier KK states…
- Present anomalies in Higgs decays like for instance on bb/ZZ, ttH and ttW.
- Invisible decays, dark matter

To conclude, assuming that the observed resonance is confirmed, the next task is to answer the usual question: 'who has ordered that?' either by embedding it in a vast ensemble like RS or by relating it to a key mechanism needed to understand, e.g., baryogenesis or dark matter. This note has tried to show that a LC operating as an e+e- collider at 750 GeV or as a photon collider using an e-e- collider at ~1 TeV, could play a crucial role to answer some of these questions.

## Acknowledgements

It is with great pleasure that I acknowledge encouragements and useful discussions with my Orsay colleagues, in particular Andrei Angelescu, Abdelhak Djouadi, Louis and Lydia Fayard, Jean-François Grivaz, Emi Kou, François LeDiberder, Yann Mambrini, Gregory Moreau, Roman Poeschl, Achille Stocchi, Dirk Zerwas and Fabian Zomer.

I am particularly grateful to Adam Falkowski for help on the graviton issue and to Michael Peskin for a careful and critical reading of this note.